\documentclass[11pt,twoside]{article}

\usepackage{asp2004}
\usepackage{psfig}
\usepackage{epsf}
\usepackage{graphics}
\usepackage{lscape}

\begin{document}

\title{X-ray Variability of AGN and the Flare Model}

\author{R.W. Goosmann$^1$, B. Czerny$^2$, A.-M. Dumont$^1$,
M. Mouchet$^1$,\\ and A. R\'o\.za\'nska$^2$}
\affil{$^1$ LUTH, Observatoire de Paris, Meudon, France \\
$^2$ Nicolaus Copernicus Astronomical Center, Warsaw, Poland}

\begin{abstract}
Short-term variability of X-ray continuum spectra has been
reported for several Active Galactic Nuclei. Significant X-ray
flux variations are observed within time scales down to
$10^3-10^5$ seconds. We discuss short variability time scales in
the frame of the X-ray flare model, which assumes the release of a
large hard X-ray flux above a small portion of the accretion disk.
The resulting observed X-ray spectrum is composed of the primary
radiation and of a reprocessed Compton reflection component that
we model with numerical radiative transfer simulations. The
incident hard X-rays of the flare will heat up the atmosphere of
the accretion disk and hence induce thermal expansion. Eventually,
the flare source will be surrounded by an optically -thick medium,
which should modify the observed spectra.
\end{abstract}

\thispagestyle{plain}

\section{Introduction}

In the last few years several Active Galactic Nuclei (AGN) showing
short-term X-ray variability have been reported (see for instance Ark
564 and Ton S180 \citep{Edelson2002}, I Zw 1 \citep{Gallo2004}, MGC-6-30-15
\citep{Vaughan2004,Ponti2004}). Modeling this behavior
remains a complex task: owing to the short variability time scales
observed, one has to assume small emission regions down to the size of
a few $R_{Schw}$ (Schwarzschild radii). Such regions exist within the
framework of the flare model. In this model compact hard X-ray sources
resulting from magnetic reconnection evolve above the accretion disk.

In the following we present detailed numerical modeling of the
reprocessed X-ray spectra induced by a magnetic flare. We investigate
the influence of a realistic flare illumination as compared to an
isotropic one and the dependency on the disk inclination. Using this
model we attempt to derive the consequences for the time evolution of
the observed spectra.

Then we briefly discuss the hydrostatic balance and the optical
properties of a the disk atmosphere below a flare.

\section{The composite spectrum of a flare-irradiated hot-spot}

\subsection{Modeling}
We assume a plane-parallel model atmosphere at a typical distance of
$9R_{Schw}$ from the central black hole. \citet{Collin2003} have shown
that, after the onset of a flare, the thermal and the ionization
profile change much quicker than the hydrostatic equilibrium. We
therefore assume the steady density structure of a non-irradiated disk
in hydrostatic equilibrium calculated according to
\citet{Rozanska1999}. The incident flux follows a power law with
$F_{inc} \propto E^{-\alpha}$, $\alpha = 0.9$ over 0.1 eV up to 100
keV. We choose a ratio $F_{inc}/F_{disk}=144$ of incident flux of the
flare to the flux coming from the disk, hence the flare by far
dominates over the disk emission.

\begin{figure}[h]
\begin{center}
\epsfxsize=6.0cm
\epsfbox{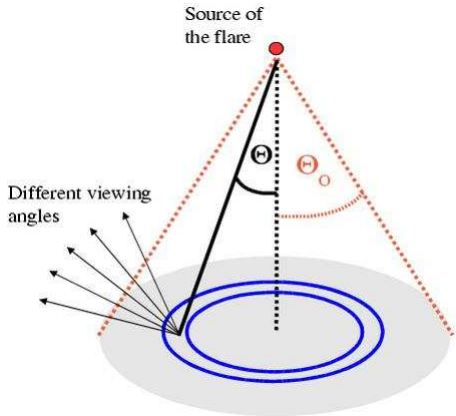}
\end{center}
\caption{Flare geometry: the observed spectrum depends on the
incident angle $\Theta$ of the irradiating photons and on the viewing
angle. $\Theta_0$ denotes the half-opening angle of the flare-cone.}
\label{fig1}
\end{figure}

In previous flare models the disk illumination was defined as
semi-isotropic \citep{Czerny2004,Collin2003} or with a constant
incident angle \citep{Nayakshin2001}.  However, the incident angle and
the ionization parameter $\xi=\frac{4 \pi F_{inc}}{n_H}$ of the flare
(with $F_{inc}$ being the incident flux and $n_H$ being the hydrogen
density at the disk surface) change significantly with the distance to
the center of the illuminated spot (Fig.\ref{fig1}). In particular we
have $\xi=\xi_0(cos\Theta)^{-3}$ if $\xi_0$ denotes the ionization
parameter at the spot center.

We have to define a maximum radius of the irradiated spot, i.e., a
maximum half-opening angle $\Theta_0$ of the flare cone. We choose
$\Theta_0=60^\circ$ because within the corresponding spot $50\%$
of the luminosity reaching the disk is reprocessed. The observed
reprocessed spectrum then consists of the integrated spectra of
successive concentric rings at incident angles with
$0^\circ<\Theta<60^\circ$. These spectra can be computed for different
inclination angles $i$ (measured from the disk symmetry axis).

We perform detailed radiative transfer simulations coupling the codes
TITAN and NOAR. TITAN simulates radiative transfer inside optically
thick media \citep{Dumont2000,Dumont2003} determining
the temperature and ionization structure. NOAR is a Monte-Carlo code
treating the Compton processes.

\subsection{Results}
\begin{figure}[t]
\begin{center}
\epsfxsize=8.2cm
\epsfbox{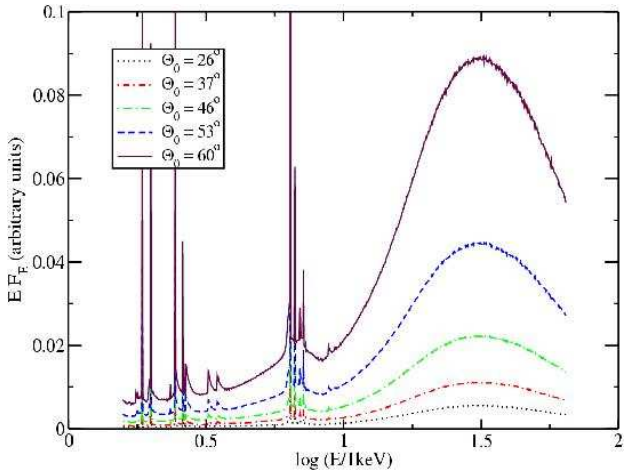}
\end{center}
\caption{Integrated reflection spectra for different half-opening
angles $\Theta_0$ seen at a typical Seyfert-1 inclination $i \sim
25^\circ$; for larger half-opening angles the spectra become harder.}
\label{fig2}
\end{figure}

The reflected spectra have been computed for successive rings of the
illuminated hot spot up to $\Theta_0=60^\circ$. While the Compton hump
in the hard X-ray range is rather similar for all incident angles
considered, the spectral slope in the soft X-rays rises for emission
regions farther away from the spot center. The integrated Compton
reflection for several half-opening angles seen at an inclination of
about $25^\circ$, which represents a typical Seyfert-1 inclination,
are shown in Fig.\ref{fig2}.

Due to different distances to the flare source the reprocessing will
evolve in time from the center of the spot toward the outer
regions. This can be seen as successively growing half-opening angles
of the flare cone. The spectra should become harder as the flare
intensifies. The time scale of this evolution is determined by the
height of the source above the disk. Such a development was observed
for the Narrow Line Seyfert 1 galaxy I Zw 1 \citep{Gallo2004}.

\section{Flare-induced evolution of the disk atmosphere}

After the dynamical time scale, the atmosphere settles into a new
hydrostatic equilibrium with a total height $H_{tot}$ and a Thomson
optical depth $\tau_{es}$. We constrain the effect of a flare on the
total height $H_{tot}$ and the Thomson optical depth $\tau_{es}$.

We analytically recompute the new equilibrium and again include the
angular dependence of the incident flux for different distances $d$
from the spot center. We find the general result that the expansion of
the disk atmosphere induced by the flare and its optical depth depend
strongly on the distance of the flare from the central black hole.

\begin{figure}[t]
\begin{center}
\epsfxsize=6.4cm
\epsfbox{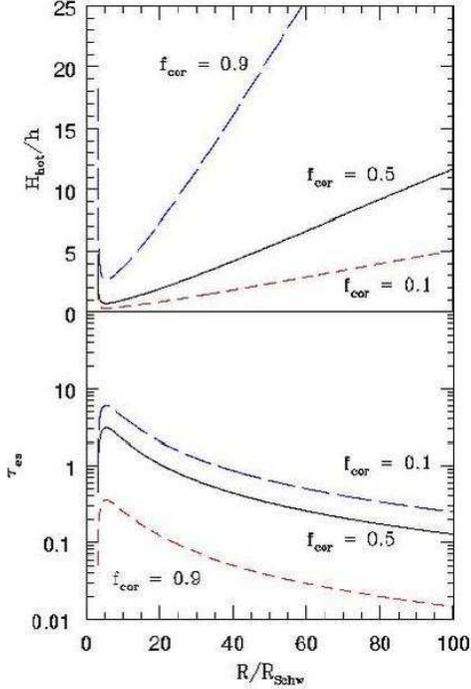}
\end{center}
\caption{Top: radial dependence of the thickness (in flare height
units). Bottom: Optical depth of the heated zone at the spot
center. The different curves denote three values of the energy
fraction $f_{cor}$ dissipated in the corona.}
\label{fig3}
\end{figure}

In Fig.\ref{fig3} we show the ratio $H_{tot}/h$, with $h$ being the
height of the flare source, and $\tau_{es}$ versus $R/R_{Schw}$. We
choose a black hole mass of $M = 10^8 M_{\odot}$, an accretion rate
$\dot m = 0.03$ (Eddington units), and a covering factor of the disk
surface with hot-spots of $f_{cover} = 0.1$ (see details of this work
in Czerny \& Goosmann \citetext{submitted}).The different curves refer
to three values of the fraction of flux dissipated in the corona. The
tendency in all cases is that toward the outer regions of the disk the
atmosphere reaches a higher altitude and becomes optically
thinner. For intermediate distances the flare source eventually will
be shrouded by an optically thick ($\tau_{es}>1$) disk atmosphere
($H_{tot}/h>1)$ which should modify the observed spectrum.

\acknowledgements{This research was supported by the CNRS No.16
``Astronomie France-Pologne'' and by the Hans-B\"ockler-Stiftung.}

\end{document}